\documentclass[aps,pre,reprint,amsmath,amssymb,superscriptaddress,showpacs,floatfix]{revtex4-1}
\usepackage{graphicx}
\usepackage{dcolumn}
\usepackage{soul,xcolor}
\usepackage[normalem]{ulem}
\usepackage{amsmath}
\usepackage{calc}
\usepackage{bm}
\usepackage[colorlinks, allcolors=blue]{hyperref}
\usepackage{setspace}
\usepackage{scalerel}
\usepackage{stackengine}
\usepackage{comment}
\stackMath
\newsavebox\tmpbox

\begin{document}

\title{Quantum phase transition in skewed ladders: an entanglement entropy \\
and fidelity study}
\author{Sambunath Das}
\affiliation{Solid State and Structural Chemistry Unit, Indian Institute of Science, Bangalore 560012, India}
\affiliation{S. N. Bose National Centre for Basic Sciences, Block - JD, Sector - III, Salt Lake, Kolkata - 700106, India}

\author{Dayasindhu Dey}
\affiliation{Solid State and Structural Chemistry Unit, Indian Institute of Science, Bangalore 560012, India}
\affiliation{UGC-DAE Consortium for Scientific Research, University Campus, Khandwa Road, Indore 452001, India}

\author{S. Ramasesha}
\affiliation{Solid State and Structural Chemistry Unit, Indian Institute of Science, Bangalore 560012, India}

\author{Manoranjan Kumar}
\affiliation{S. N. Bose National Centre for Basic Sciences, Block - JD, Sector - III, Salt Lake, Kolkata - 700106, India}


\begin{abstract}
	Entanglement entropy (EE) of a state is a measure  of correlation or entanglement between two parts of a composite system and
	it may show appreciable change when the ground state (GS) undergoes a qualitative change in a quantum phase transition (QPT). 
	Therefore, the EE has been extensively used to characterise the QPT in various correlated Hamiltonians. Similarly fidelity also 
	shows sharp changes at a QPT. We characterized the QPT of frustrated antiferromagnetic Heisenberg spin-1/2 systems on 3/4, 3/5 
	and 5/7 skewed ladders using the EE and fidelity analysis. It is noted that all the non-magnetic to magnetic QPT boundary in 
	these systems can be accurately determined using the EE and fidelity, and the EE exhibits a discontinuous change, whereas 
	fidelity shows a sharp dip at the transition points. It is also noted that in case of the degenerate GS, the unsymmetrized 
	calculations show wild fluctuations in the EE and fidelity even without actual phase transition, however, this problem is  
	resolved  by calculating the EE and the fidelity in the lowest energy state of the symmetry subspaces, to which the degenerate 
	states belong.
\end{abstract}

\maketitle

\section{\label{sec:intro}Introduction}
Quantum phase transition (QPT) involves a qualitative change in the nature of the ground state (GS) in the phase space of the Hamiltonian
parameters. The prerequisite for a Hamiltonian to exhibit QPT is that it should consist of noncommuting terms. In this case the
quantum fluctuations drive the QPT. In 1-D and 2-D Hamiltonians the fluctuations are dominant, making them viable candidates to exhibit
QPT. Many interesting and exotic quantum phases observed include the dimer
phase ~\cite{Majumdar,Chubukov,Chitra,White,Itoi,Mahdavifar,Sirker,M_Kumar1,Z_Soos,M_kumar2}, spin liquid
phase ~\cite{Hamada,White,Itoi,Mahdavifar,M_Kumar1,Z_Soos}, charge density waves ~\cite{Manoranjan1,Manoranjan2,Hirsch1,Hirsch2}, spin
density waves ~\cite{Daya_SDW,White_Huse,Sorensen_Affleck}, vector chiral phase~\cite{Chubukov,Hikihara,Sudan,Aslam}, the valence
bond solid~\cite{Affleck1,Affleck2,Schollwock} and topological phases ~\cite{Tp_Gu, Tp_pollmann,Haldane1,Haldane2}, to name a few. The QPT is different from the classical or thermal phase transition in which phase transition is driven by the competition between energy and thermal
entropy at finite temperature. Quantum fluctuations are dominant in confined systems like one, quasi-one or two dimensional systems and
lead to many interesting quantum phases. In a quantum system the EE which indirectly measures the correlation between  different
parts of the system is expected to change in QPT. The extent of entanglement depends on Hamiltonian parameters such as system geometry,
competing interaction parameters, magnetic and electric fields.

Effective low dimensional systems such as spin chains and spin ladders are of enduring interest for the past many decades. Fundamentally
different spectral gaps in spin-1/2 and spin-1 chains predicted by Haldane, spin-Peierls’ distortion in spin-1/2 chains predicted from
theory, and occurrence of magnetization plataeus in spin ladders have all been observed experimentally ~\cite{Kikuchi1,Kikuchi2,Gu,Hase,Maignan,Hardey1,Hardey2,Ishiwata,wang,Yao,Lenertz,He,Shiramura}. Thus, most theoretical predictions of spin chains have been vindicated by 
experiments, mostly in molecular systems and magnetically low-dimensional systems characterized by strong exchange in one direction and 
much weaker exchange in orthogonal directions  \cite{Dagotto}. In fact, experimentally, there is evidence of systems  \cite{Castilla} which 
belong to different regions of the quantum phase diagram in the well studied $J_1$ – $J_2$ – $\delta$  model of spin chains, where $J_1$ is 
the first neighbour exchange interaction, $J_2$ is the second neighbour exchange interaction and $\delta$ is the dimerization in the 
nearest neighbour exchange strength.

Belonging to the class of spin ladders is a class of skewed ladders where one or more rung bonds in the unit cell is slanted. This ladder
system can also be mapped into zigzag ladders with periodically missing bonds as shown in Fig.~\ref{fig:schematic}. The skewed ladder
is classified as \textit{n/m} ladder where \textit{n} and \textit{m} are the ring sizes of adjacent rings in the ladder. These systems, in 
our opinion, are realizable in concatenated transition metal complexes. This class of systems show a very interesting quantum phase diagram 
as a function of the ratio of the rung exchange ($J_1$) to the nearest neighbour leg exchange ($J_2$). Depending upon this ratio \textit{n/m} and ${J_1}/{J_2}$ and system size, the skewed ladders show remarkable switching in the GS spin, exhibit bond order wave, spin density wave 
and chiral vector phases \cite{Geet}. The Heisenberg antiferromagnetic (HAF) spin-1/2 system on 5/7, 3/4, 3/5 and 5/5 skewed ladders shows interesting magnetic and non-magnetic QPT in the parameter space of the rung and the leg spin exchange constants \cite{Geet,Daya}. Recent work 
on HAF spin-1 on 5/7 ~\cite{sambu_57_spn1}, 3/4 and 3/5 \cite{sambu_3435} skewed ladders has also shown existence of similar magnetic and non-magnetic phase transitions. The quantum phases of HAF spin-1/2 system on 3/4, 5/5 and 3/5 skewed ladders studied at finite external axial 
magnetic field $B$ are found to show various magnetic plateau phases which follow the Oshikawa, Yamanaka, and Affleck (OYA) \cite{OYA} rule, 
in the $B-J_1-J_2$ parameter space.

In general correlation functions, energy crossovers, GS symmetry changes and local order parameters are studied to characterize the QPT.
However, hidden order like topological properties can not be detected by conventional procedures. Therefore, to study the long range correlations in the system and to characterize the phase boundaries sharply, we have studied the entanglement properties of this system. Where
entanglement does not show sharp features, fidelity of the GS is a useful tool for determining the phase boundary. The QPT of
frustrated $J_1-J_2$ model has been characterised using EE in earlier studies \cite{Li_EE}. Simillarly the QPT of XXZ model has
been studied using the fidelity as a property obtained from exact diagonalization and density matrix renormalization group
methods \cite{Luo_XXZ}. The fidelity approach to characterize QPT is reviewed in \cite{GU_JIAN}.

We find that fidelity is very sensitive to changes in the GS and shows sharp changes even when the spin gap vs $J_1/J_2$ plots show only a 
change in the slope or when there is a cross over in the two low-lying excited states. When there is symmetry breaking, the fidelity in 
one of the subspaces shows a sharp change while in another subspace there is no change. This allows us to identify the symmetry before and 
after a quantum phase transition. In this work, we follow the changes in EE and fidelity of different skewed spin-1/2 ladders. We mainly 
focus on exact diagonalization studies of these ladders. We also employ symmetries such as the conservation of the z-component of total spin, $S_z$, and the reflection symmetry at degeneracies to follow changes in EE and fidelity. We find that in the 3/4 skewed ladder, EE shows 
sharp changes at the critical value of $J_{1}$. Fidelity also shows a similar sharp changes and we can identify the critical $J_{1}$ value
very accurately. In the skewed ladders 3/5 and 5/7, both EE and fidelity show sharp changes and the critical $J_{1}$ values for QPT can be 
determined highly accurately. In what follows, we discuss the results of our studies on the 3/4, 3/5 and 5/7 ladder systems and show how both
EE and fidelity can be used to accurately find critical $J_{1}$ values at the transition points.

This paper is divided into five sections. In section~\ref{sec2} we provide a brief introduction to entanglement entropy and fidelity.
We then present the model Hamiltonian and numerical methods in section~\ref{sec3}. The results are discussed in section~\ref{sec4} under 
three subsections. The summary of our results are presented in section ~\ref{sec5}.
\section{\label{sec2}Entanglement entropy and Fidelity}
Entanglement entropy (EE) has been studied extensively from a quantum information perspective due to the promise that quantum computing
holds. It has also been shown in many studies that EE is a useful tool for studying phase transition in quantum systems. Given two
subsystems A and B and spanning Hilbert spaces $H_A$ and $H_B$ with full system Hilbert space
$H = H_{A} \times H_{B}$, we can always
find an orthonormal bases $\{\phi_A\}$ and $\{\phi_B\}$ such that any state $\vert{\psi}_{AB}\rangle$ in the Hilbert space belonging to $H$,
can be expressed as
\begin{equation}
 \vert\psi_{AB}\rangle = \sum_{i,j}  \alpha_{ij}  \vert\phi_{i}\rangle_{A}  \vert\phi_{j}\rangle_{B}.
\end{equation}
The reduced density matrix of the subsystem B is given by
\begin{equation}
	\rho_{B} = Tr_{A} \left(\vert\psi_{AB}\rangle \langle\psi_{AB}\vert\right),
\end{equation}
and the $jj^{'th}$ matrix element of the reduced density matrix $\rho_B$ is, 
\begin{equation}
	{(\rho_{B})}_{jj^{'}} = \sum_{i} \alpha_{ij} \alpha{^{*}}_{ij{^{'}}}.
\end{equation}
If the eigenvalues of $\rho_{B}$ are $\lambda_{i}$ then we can define EE in many different ways. If we restrict ourselves to extensivity
of entropy, i.e. $S_{AB} = S_{A} + S_{B}$, we can define entropy of a state in two ways. In condensed matter physics, the widely used 
entropy is the von Neumann entropy which is given by~\cite{Petz2001}
\begin{equation}
S = -\sum_{i} \lambda_{i} \log_{2} \lambda_{i},
\label{eq:EE}
\end{equation}
where $\lambda_{i}^{\prime}$s are the eigenvalues of the reduced density matrix. Von Neumann entropy is the generalization of the classical
Shannon entropy. Another entropy function which is R\'enyi entropy $S_{A,\alpha}^{R}$ is defined as~\cite{Renyi}
\begin{equation}
        S_{A,\alpha}^{R} = \frac {1}{1-\alpha} \textrm{log} \left [\sum_{i} (\rho_{i}^{A})^{\alpha} \right]
\end{equation}
where $\alpha$ can take values between 0 and $\infty$. R\'enyi entropy reduces to von Neumann entropy in the limit $\alpha \to 1$ and it 
is maximum when $\alpha = 0$ and minimum when $\alpha \to \infty$.

In this work, we employ von Neumann entropy to analyze the QPT as von Neumann entropy is the most widely used entropy to study the
QPT. We also restrict ourselves to equal sizes of the subsystems A and B as von Neumann entropy is maximum in this case.
In frustrated one-dimensional spin chains, which are known as the $J_{1}-J_{2}$ chains, the GS EE cannot identify the critical $J_2$ value 
($J_{2c}$) below which the system is gapless and above which the system is gapped. However, the GS EE can identify the Majumdar-Ghosh point 
which is not surprising since the GS corresponds to nearest neighbor singlets ~\cite{Durga}. The $J_{2c}$ is correctly identified from EE 
when the $J_{2}$ corresponding to crossing of the EE of the first excited triplet and the second singlet is extrapolated from finite size 
calculations to the thermodynamic limit ~\cite{Durga}. EE has also been studied in spin ladders and it has been shown that the area law of 
entanglement is valid upto seven legs~\cite{Kallin}.

A more sensitive property to follow QPT is fidelity and fidelity susceptibility. Fidelity ($F(\omega)$) at a parameter $\omega$ of the
Hamiltonian is given by the overlap of the desired state of the Hamiltonian at $\omega$ with that in its neighborhood $\omega+\delta\omega$ i.e.
\begin{equation}
        F(\omega) = \langle \psi(\omega)\vert\psi(\omega+\delta\omega)\rangle.
\label{eq:Fidelity}
\end{equation}
We can also define Fidelity susceptibility $\chi(\omega)$ as
\begin{equation}
 \chi(\omega) = \frac{2(1-F(\omega))}{(\delta\omega)^2}
\end{equation}
which follows from a Taylor series expansion of $F(\omega)$. It has been shown that fidelity of the GS in the $J_{1}-J_{2}$ model can give 
accurate value of $J_{2c}$ ~\cite{Thesberg}. In the $J_{1}-J_{2}$ model, the critical value of $J_{2}$ has also been determined by following 
the fidelity of the first excited state of short Heisenberg chains to determine $J_{2}(N)$ and extrapolating it to the thermodynamic limit 
$N\to\infty$ ~\cite{Chen}. In the $J_{1}-J_{2}-\delta$ model, using EE, it is not possible to obtain the quantum phase diagram accurately; 
the EE contours show phase changes although phase boundries remain fuzzy. We have obtained the EE and fidelity of the 3/4, 3/5, and 5/7 
skewed ladders; these systems show quantum phase transition for fixed next nearest neighbor exchange $J_2$ and varying nearest neighbor 
$J_1$ exchange.
\section{\label{sec3}Model Hamiltonian and Numerical method}
\begin{figure}[ht!]
\begin{center}
\includegraphics[width=2.9 in]{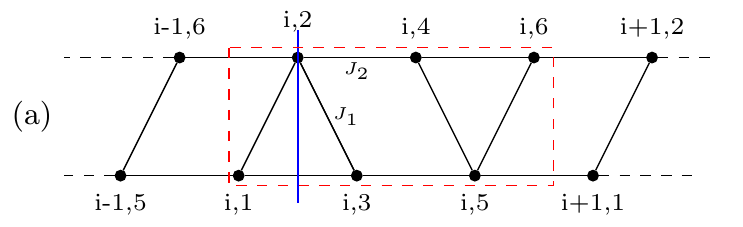}\\
\vspace{0.6cm}
\includegraphics[width=2.9 in]{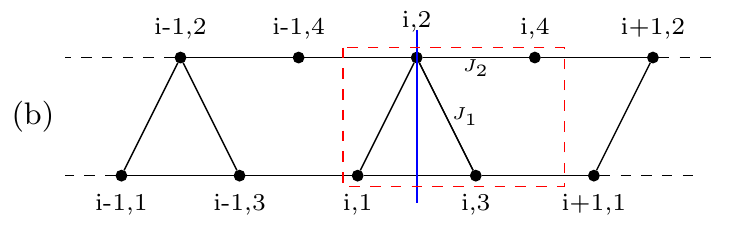}\\
\vspace{0.6cm}
\includegraphics[width=2.9 in]{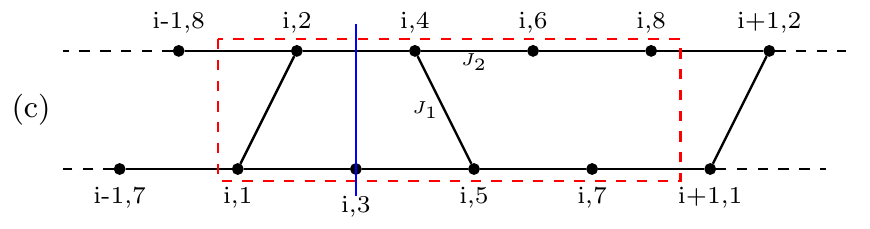}
\caption{\label{fig:schematic}Schematic diagram of (a) 3/4 skewed ladder, (b) 3/5 skewed ladder, and
        (c) 5/7 skewed ladder. The nearest neighbor or rung interaction is $J_1$ and the next-nearest neighbor
        (according to our numbering scheme) interaction is $J_2$. The sites on the top leg are even numbered and on
	the bottom leg are odd numbered. The unit cell for each ladder system is indicated by a rectangle with dashed red lines.
	The mirror plane for each ladder system is represented by a perpendicular blue line.}
\end{center}
\end{figure}

In our study, we consider a spin-1/2 model on 3/4, 3/5 and 5/7 skewed ladders as shown in Fig.~\ref{fig:schematic}(a), (b) and (c).
The rung bond interaction is $J_1$ and the leg bond interaction is $J_2$. All the exchange interactions in the Hamiltonians we consider are
antiferromagnetic in nature. The sites are numbered in such a way that the even numbered sites are on the top of the leg and odd numbers 
are on the bottom of the leg. The leg bond interaction $J_2$ is set to 1 and it defines the energy scale.
The model Hamiltonian of 3/4 skewed ladder is written as
\begin{eqnarray}
	H_{3/4} &=& J_1 \sum_i \left[\left(\vec{S}_{i,1} + \vec{S}_{i,3}\right)\cdot \vec{S}_{i,2} +
        \left(\vec{S}_{i,4} + \vec{S}_{i,6} \right) \cdot \vec{S}_{i,5} \right]  \nonumber \\
        & &+ J_2 \sum_i \bigg( \vec{S}_{i,5} \cdot
        \vec{S}_{i+1,1}
        + \vec{S}_{i,6} \cdot \vec{S}_{i+1,2}  \nonumber \\
        & &  \qquad \qquad + \sum_{k=1}^4
        \vec{S}_{i,k} \cdot \vec{S}_{i,k+2} \bigg)
\label{eq:ham34}
\end{eqnarray}
where $i$ labels the unit cell and $k$ the spins within the unit cell. The first term denotes the rung bond interaction and the
second term denotes the leg bond interaction. Similarly, the model Hamiltonian for the 3/5 and 5/7 systems can be written as
\begin{eqnarray}
	H_{3/5} &=& J_1 \sum_i \left(\vec{S}_{i,1} + \vec{S}_{i,3}\right)\cdot \vec{S}_{i,2}
        + J_2 \sum_i \bigg( \vec{S}_{i,3} \cdot \vec{S}_{i+1,1} \nonumber \\
        & &+ \vec{S}_{i,4} \cdot \vec{S}_{i+1,2}
        + \sum_{k=1}^2 \vec{S}_{i,k} \cdot \vec{S}_{i,k+2} \bigg),
\label{eq:ham35}
\end{eqnarray}
and
\begin{eqnarray}
H_{5/7} &=& J_1 \sum_i \left(\vec{S}_{i,1} \cdot \vec{S}_{i,2} + \vec{S}_{i,4}
        \cdot \vec{S}_{i,5} \right)  \nonumber \\
        & &+ J_2 \sum_i \bigg( \vec{S}_{i,7} \cdot
        \vec{S}_{i+1,1}
        + \vec{S}_{i,8} \cdot \vec{S}_{i+1,2}  \nonumber \\
        & &  \qquad \qquad + \sum_{k=1}^6
        \vec{S}_{i,k} \cdot \vec{S}_{i,k+2} \bigg).
\label{eq:ham57}
\end{eqnarray}
We have carried out the computations using exact
diagonalization technique on a system with 24 spin-1/2 sites, in all cases. We have employed cyclic boundary conditions and the EE
computed is the von Neumann entropy. The environment and the system, both have equal number of sites, i.e. 12 sites.
The reduced density matrix of the left half of the system is obtained by tracing over the right half of the system,
\begin{equation}
 \rho_{L{L^{'}}} = \sum_{R} {C_{LR}} C_{L^{'}R}
\end{equation}
where the GS wavefunction $\psi_{g}$ is expressed in the direct product basis of the Fock space of the system ($L$) and the
surroundings ($R$). The dimensionality of the Fock space of the system  block, and hence the order of the density matrix is
$2^{12}\times 2^{12}$ or $4096\times 4096$. After obtaining all the eigenvalues $\{\lambda_{i}\}$ of the reduced density matrix, the 
von Neumann entropy is obtained from Eqn.~\ref{eq:EE}. We have computed the fidelity $F(J_1)$ of the state using the normalized ground 
states and Eqn.~\ref{eq:Fidelity}. We have varied $J_{1}$ in very small steps of $\delta(=0.001)$ near the quantum phase transitions 
obtained from the spin gap data. The spin gap $\Gamma_l$ is defined as the energy gap between the lowest eigenstates $S_z = l$ and
$S_z = 0$ manifolds, where $S_z$ is the z-component of the total spin.
\begin{equation}
        \Gamma_l = E_{0} (S_{z}=l) - E_{0} (S_{z}=0),
\end{equation}
where $l$ is an integer. The GS spin $(S_G)$ $= l$, for $l$ that satisfies $\Gamma_l = 0$ and $\Gamma_{l+1} > 0$.
The excitation gap $\Gamma_{\sigma}$ at a fixed $S_G$ between the lowest energy states in different reflection symmetry subspaces
$\sigma(-)$ and $\sigma(+)$ is defined as
\begin{equation}
       \Gamma_{\sigma} = E_{0} (S_G,  \sigma(-)) - E_{0} (S_G,  \sigma(+)).
\end{equation}
The GS is in odd subspace ($\sigma(-)$) when $\Gamma_{\sigma}< 0$, even subspace ($\sigma(+)$) when $\Gamma_{\sigma}> 0$ and doubly degenerate
when $\Gamma_{\sigma}= 0$.
\section{\label{sec4}Results and discussions}
\subsection{3/4 skewed ladder}
\begin{figure}
\begin{center}
\includegraphics[width=2.9 in]{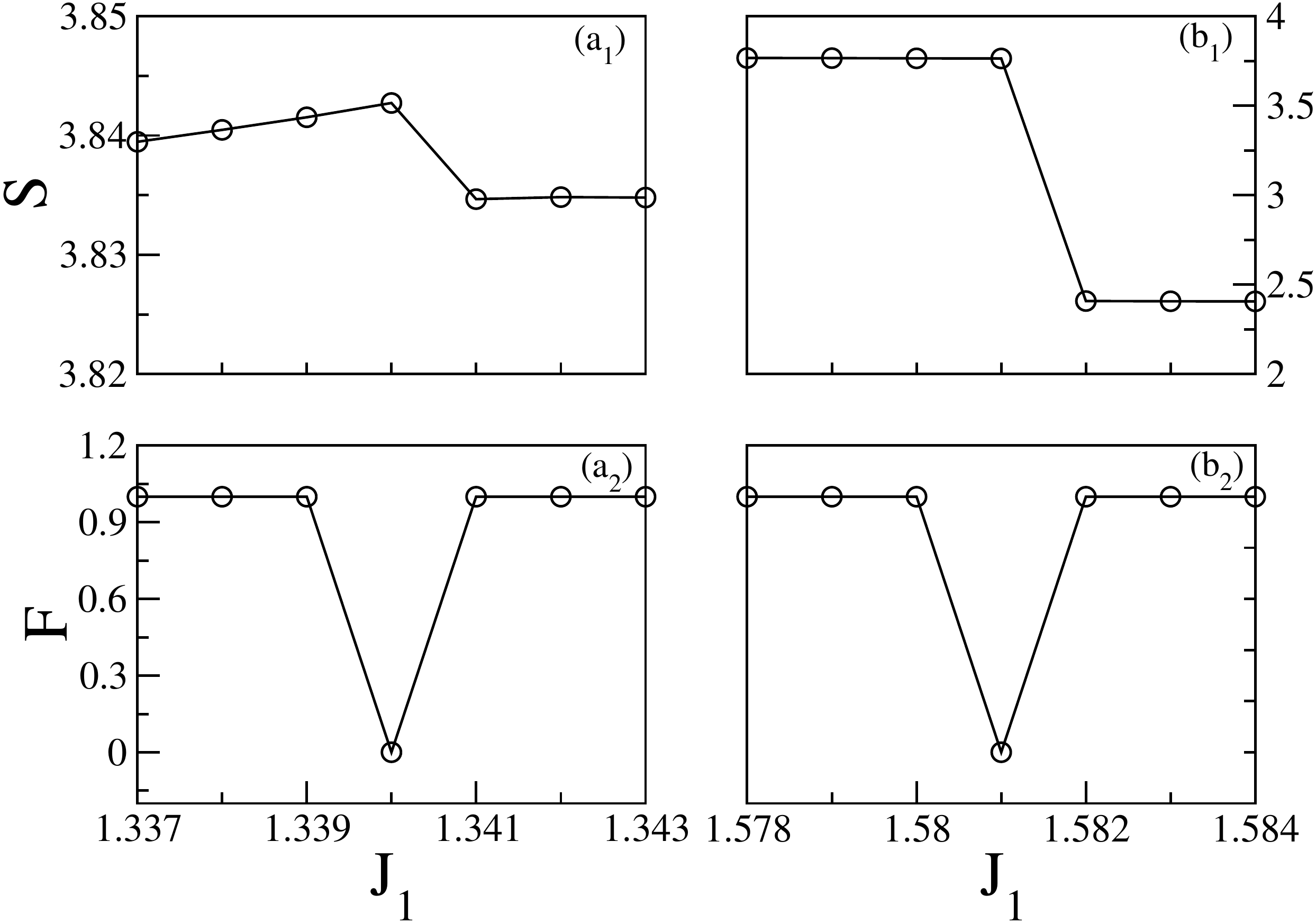}
        \caption{\label{fig:EE_fid_34} The behaviour of EE (S) and fidelity (F) in the range $1.337 \le J_1 \le 1.343$ of a
	3/4 skewed ladder with N = 24 spins are shown in ($a_1$) and ($a_2$) respectively. The EE changes abruptly and a sharp
	drop in fidelity occurs at $J_1=1.340$. At other values of $J_1$ fidelity is constant (=1). Simillary the behaviour of EE and
        fidelity in the range $1.578 \le J_1 \le 1.584$ are shown in ($b_1$) and ($b_2$). At $J_1=1.581$, the entropy changes and 
	fidelity shows a sharp drop.}
\end{center}
\end{figure}

This is the simplest of the skewed ladders to show a transition in the spin of the GS. However, there is a transition in the GS near 
$J_{1} = 1.3$ which does not involve change in the GS spin.
\begin{figure}
\begin{center}
\includegraphics[width=2.9 in]{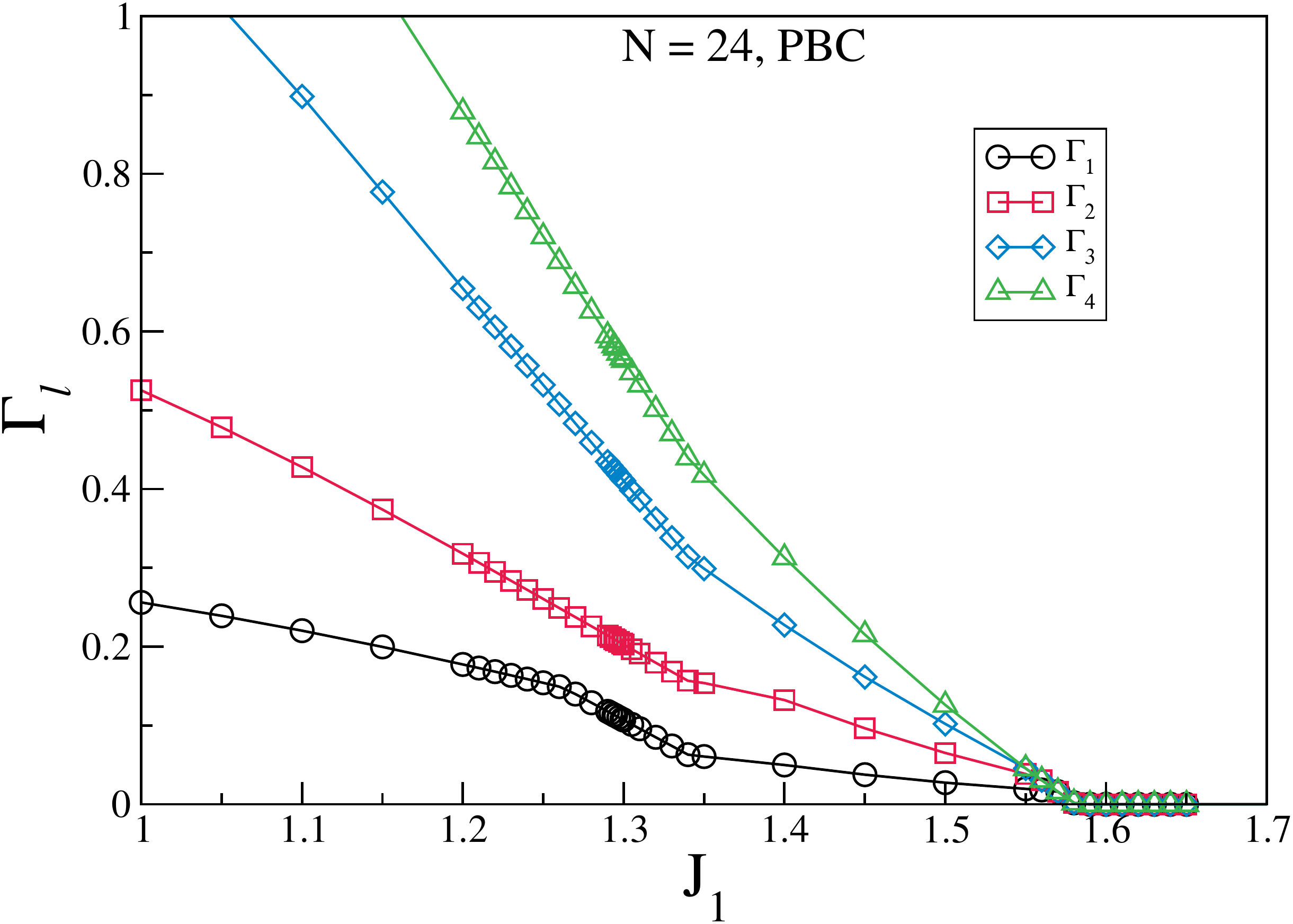}
\caption{\label{fig:7_1} Spin gaps $\Gamma_l$ for a 3/4 skewed ladder of 24 spins with PBC shown as a
	function of $J_1$. The GS spin $S_G = 0$ for $J_1<1.581$ and $S_G = 4$ for $J_1 \geq 1.581$.}
\end{center}
\end{figure}
We notice in the $\Gamma_{l}$ vs $J_1$ plot (Fig.~\ref{fig:7_1}) that there is a kink near $J_1 = 1.3$,
but it is not possible to pin point the $J_1$ value at which this kink appears from the $\Gamma_{l}$ plot. However, the EE
shows a sharp change at $J_{1} = 1.340$ and it is also seen more clearly in the fidelity, where it vanishes at $J_{1} =1.340$ but has a
value of 1.0 for other $J_1$ values (Fig.~\ref{fig:EE_fid_34}). Closer examination of the full eigenvalue spectrum of the Hamiltonian
shows that at this value of $J_1$, the singlet GS is doubly degenerate and there is a cross over from one
singlet GS to another singlet GS. The transition to the highest spin states occurs sharply, when the GS spin
changes to $S_G = 4$.
The transition occurs at $J_{1}= 1.581$ and there is a sharp drop in EE at this point. Fidelity also goes to zero at $J_{1}=1.581$ but
assumes a value of $1$, at the neighboring points. For larger system sizes, the level crossing method is used to calculate the transition
points, and it has been reported that the transition points do not change significantly with system size (ref.~\cite{Geet}). In this system, 
we have been able to determine the parameter at which the QPT occurs very accurately and we also find that there is a transition at 
$J_{1}=1.340$ which is not evident from the plots of the spin gaps.
\subsection{3/5 skewed ladder}
\begin{figure}
\begin{center}
\includegraphics[width=2.9 in]{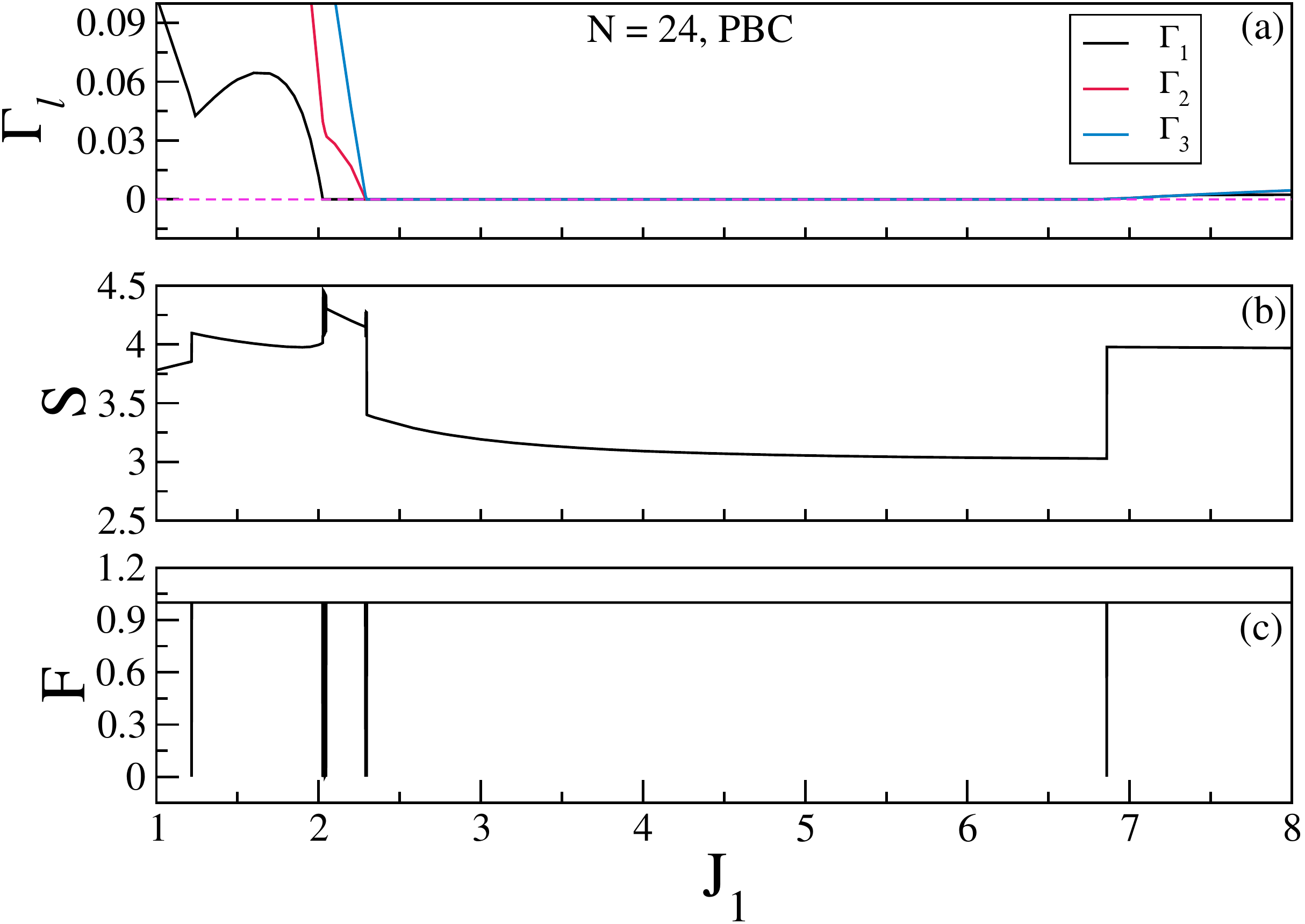}
        \caption{\label{fig:gap_35}(a) The spin gaps $\Gamma_1$, $\Gamma_2$ and $\Gamma_3$ are shown as a function of $J_1$ for N = 24
        sites in a 3/5 skewed ladder with PBC. For $J_1$ $<$ 2.026 and $J_1$ $>$ 6.859, $\Gamma_1$ becomes 0 and the system shows 
	nonmagnetic behaviour. In the region $2.026 \le J_1 \le 2.297$ the GS spin gradually changes from 0 to 3. (b) shows the EE and 
	(c) shows the fidelity for the unsymmetrized GS as a function of $J_1$. The EE exhibits a discontinuous change, while fidelity 
	shows a sharp drop at the transition points. The thick line near the vicinity of $J_1=2$ indicates many transitions both in EE and 
	fidelity.}
\end{center}
\end{figure}

\begin{figure}
\begin{center}
\includegraphics[width=2.9 in]{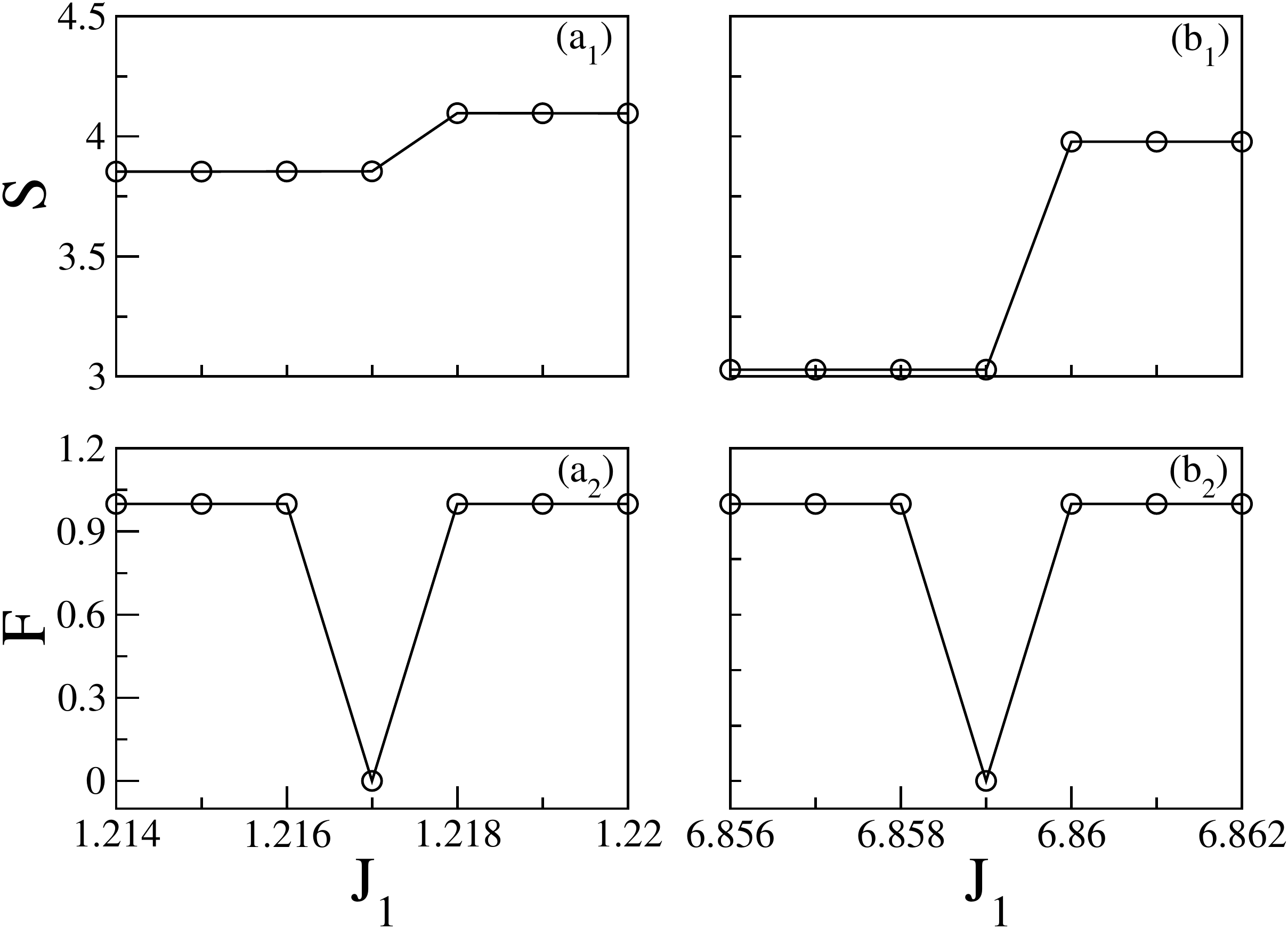}
        \caption{\label{fig:35_EE_fid_1.217_6.859} The behaviour of EE (S) and fidelity (F) in the range of $J_1$
        ($1.214 \le J_1 \le 1.22$) of a 3/5 skewed ladder with N = 24 spins are shown in ($a_1$) and ($a_2$) respectively.
        The change in entropy and sharp drop in fidelity occurs at $J_1=1.217$. For other values of $J_1$ the fidelity is 
	constant (=1). Simillary behaviour of the entropy and fidelity in the range of $J_1$ ($6.856 \le J_1 \le 6.862$) are 
	shown in ($b_1$) and ($b_2$) respectively. At $J_1=6.859$, the entropy changes and fidelity shows a sharp drop.}
\end{center}
\end{figure}

The GS of 3/5 skewed ladder is nonmagnetic for $J_{1} < 2.026$ and the spin of the GS progressively increases from
$S_{G} = 0$ to $S_{G} = 3$ form $J_{1} = 2.026$ to $2.297$ (Fig.~\ref{fig:gap_35}(a)).
The spin of the GS remains 3 as $J_1$ is increased further until $J_{1} = 6.859$ and for $J_{1} > 6.859$ the state reenters the nonmagnetic
phase. EE and fidelity are shown in (Fig.~\ref{fig:gap_35}(b),(c)) over the entire range of our study, namely $1 \le J_{1} \le 8$. 
In this ladder system, both EE and fidelity give the critical value of $J_1$ to be 1.217 (Fig.~\ref{fig:35_EE_fid_1.217_6.859}) and at this 
value of $J_1$ both the lowest energy state and first excited state are same upto third decimal space in $S_z=0$ sector. Here again it is
likely due to broken spatial symmetry in the GS. At $J_1 = 6.859$, we find a sharp transition in both EE and fidelity to 
the reentrant nonmagnetic phase.

For $J_{1}$ in the neighborhood of 2, fidelity and EE seem to indicate many transitions. The system has a reflection symmetry, 
as shown in Fig.~\ref{fig:schematic}(b). The lowest energy states for $2.026\le J_1 \le 2.046$ and  $2.290\le J_1 \le 2.296$, in 
the $\sigma(+)$ and $\sigma(-)$ manifold are degenerate. A linear combination of the degenerate states can lead to a broken symmetry 
state (see ref.~\cite{Geet} for details). The intervals of $J_1$ for which the GS is doubly degenerate are shown in third column of 
Table ~\ref{tab:35}. 
\begingroup
\begin{table}
 \begin{center}
	 \caption{\label{tab:35} The interval of $J_1$ for the GS in different subspaces of a 3/5 skewed
	 ladder with $N = 24$ spins. The first column represents the range of $J_1$ for which the GS is in
	 $\sigma(+)$ subspace, the second column represents the GS in $\sigma(-)$ subspace. The third
	 column gives the range of $J_1$ for which the difference between the lowest energies of two 
	 different subspaces is zero.}
\noindent\hrulefill
\smallskip\noindent
\resizebox{\linewidth}{!}{%
\begin{tabular}{|l|l|l|}
\hline
\multicolumn{1}{|l|}{\quad \quad \quad $\sigma(+)$}
& \multicolumn{1}{|l|}{\quad \quad \quad $\sigma(-)$}
& \multicolumn{1}{|l|}{\quad \quad \quad $|\Gamma_{\sigma}|=0$} \\ \hline
\quad $1<J_1\le 1.217$ & $1.217\le J_1 < 2.025$ & $2.026\le J_1 \le 2.046$ \\
\quad \quad \quad and & \quad \quad \quad and & \quad \quad \quad and \\
$2.046<J_1<2.290$ & \quad \quad $J_1>2.296$ & $2.290\le J_1 \le 2.296$ \\ \hline
\end{tabular}}
\end{center}
\end{table}
The EE of the eigenstate with symmetry shows a kink at $J_{1}=2.025$ below which the GS is in the $\sigma(-)$ space ($\sigma$ is the
reflection symmetry) and the EE in the $\sigma(+)$ space shows a kink at $2.046$ (Fig.~\ref{fig:with_without_35}(a)). The fidelity of the
GS in $\sigma(-)$ space show a drop at $2.025$ and that in the $\sigma(+)$ space shows a drop at 2.046 (Fig.~\ref{fig:with_without_35}(b)).
\begin{figure}
\begin{center}
\includegraphics[width=2.9 in]{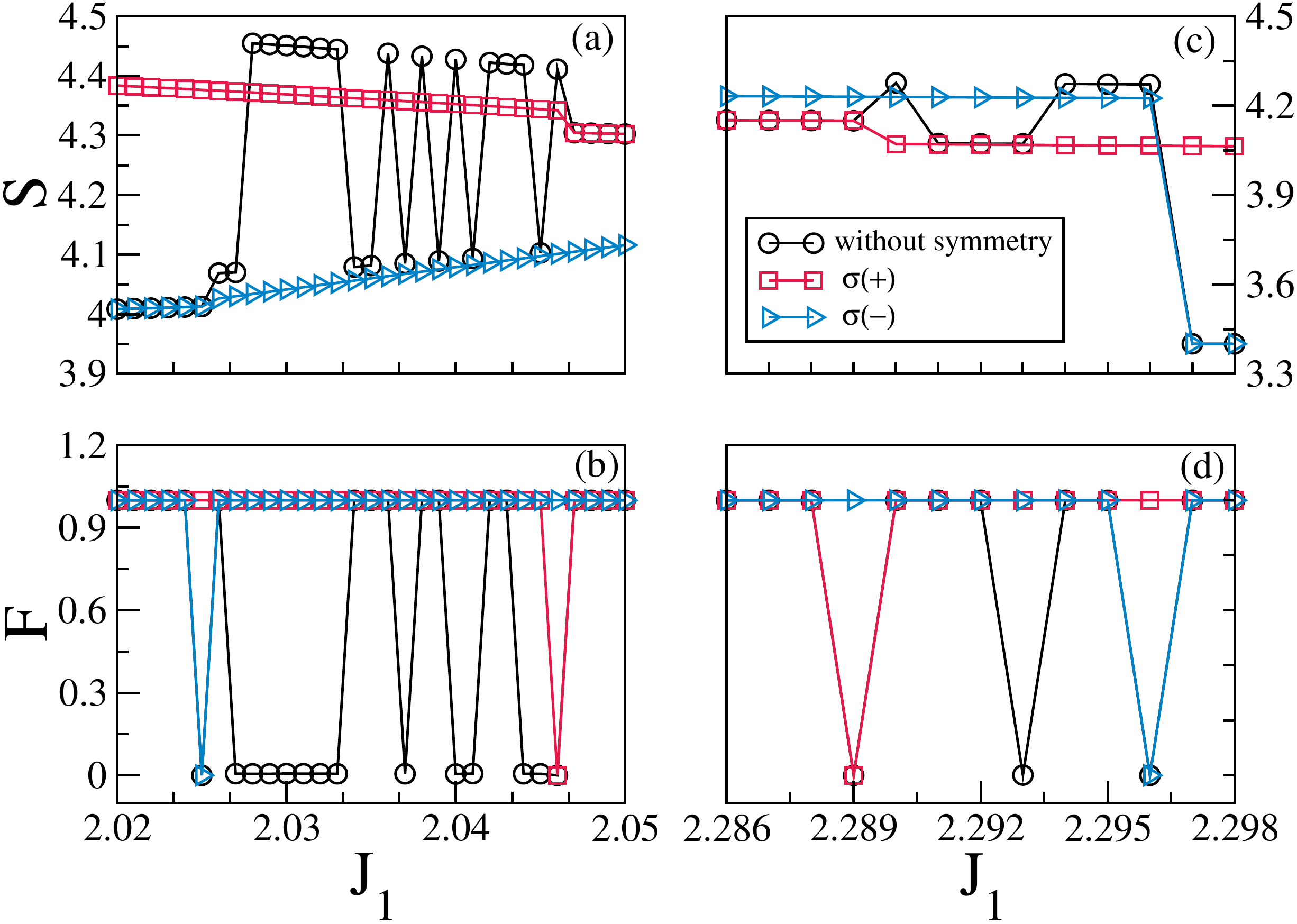}
        \caption{\label{fig:with_without_35} The EE (S) and fidelity (F) of lowest energy states of a 3/5 skewed ladder for 24 sites with
        periodic boundary condition in different reflection symmetry subspaces (($\sigma(+)$) and ($\sigma(-)$).
	Both the EE and fidelity for the unsymmetrized GS is represented by the black curve whereas for the symmetrised 
	GS these are represented by the red and blue curves.}
\end{center}
\end{figure}
In the region between these values, the GS is doubly degenerate and any linear combination of these states is also an eigenstate and hence
the EE and fidelity, calculated from unsymmetrized GS show wild fluctuations. For $2.290 \le J_1 \le 2.296$ the GS is degenerate. So the 
unsymmetrized calculation for both EE and fidelity shows wild fluctuations in this range of $J_1$. In symmetrized calculation, both the EE 
and fidelity change at $J_1=2.289$ and 2.296 (Fig.~\ref{fig:with_without_35}(c) and \ref{fig:with_without_35}(d)). The $|\Gamma_{\sigma}|$ 
value vanishes in the interval $2.290 \le J_1 \le 2.296$. The GS below 2.290 has $\sigma(+)$ symmetry whereas above 2.296 the GS has 
$\sigma(-)$ symmetry. Therefore fidelity for the $\sigma(+)$ subspace shows a sharp dip at $J_1=2.289$ and that of state $\sigma(-)$ shows 
a dip at $J_1=2.296$.
\subsection{5/7 skewed ladder}
Among the skewed ladders we have studied, the 5/7 skewed ladder has by far the most interesting quantum phase transitions. For fixed $J_2 = J_1 =1$, the GS spin of the system, $S_G$, increases with system size, a feature we have not observed in the other systems studied here. In this work, we focus on the EE and fidelity for a 24 spin 5/7 ladder with periodic boundary condition as a function of $J_1$. From the plot of spin gaps vs. $J_1$(Fig.~\ref{fig:nosym_all_57}(a)) and
\begin{figure}
\begin{center}
\includegraphics[width=2.9 in]{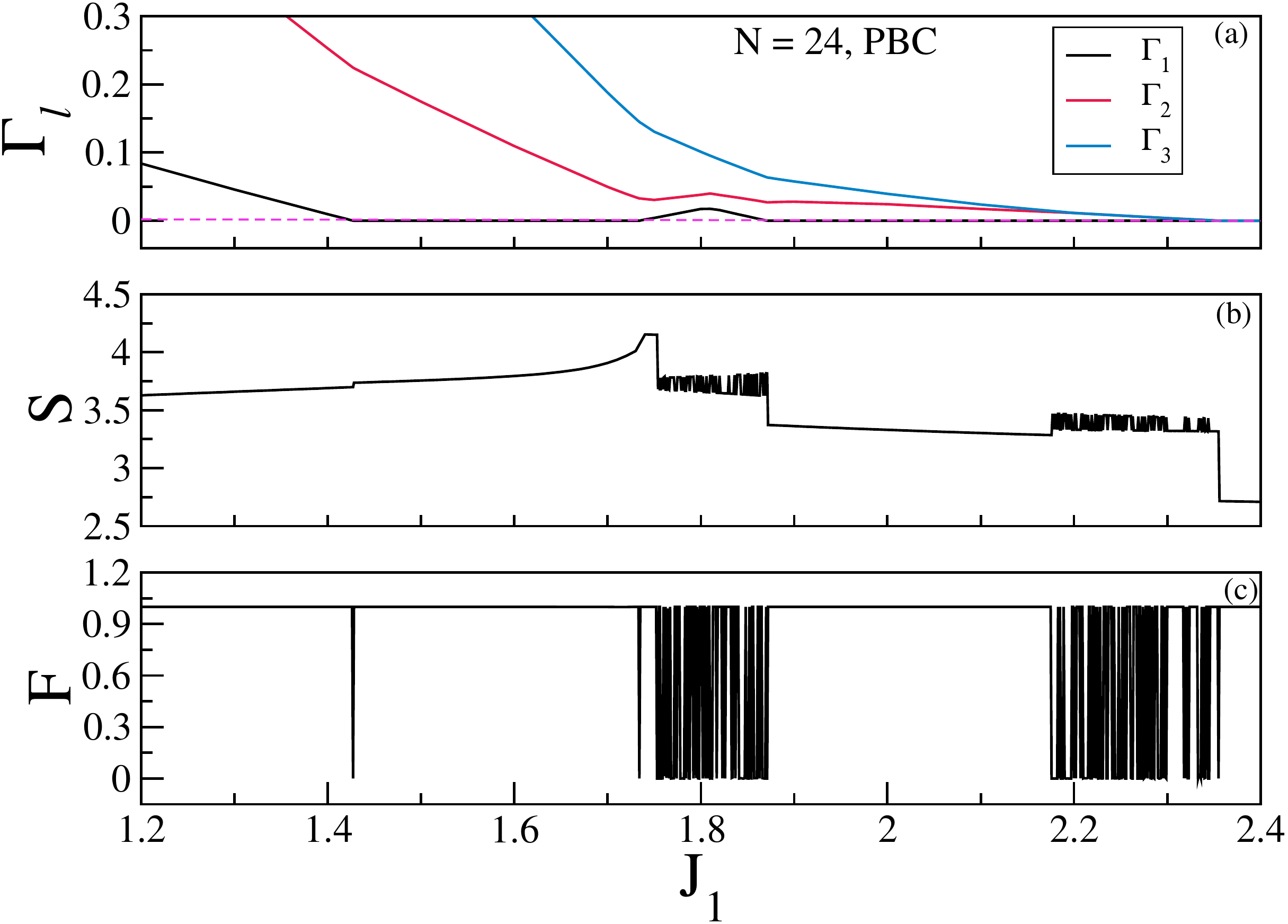}
\caption{\label{fig:nosym_all_57}(a) The spin gaps $\Gamma_1$, $\Gamma_2$ and $\Gamma_3$ are shown as a function of $J_1$ for 24
        site spins in a 5/7 skewed ladder with PBC. For $J_1 < 1.427$, $\Gamma_1$ becomes nonzero and the system shows nonmagnetic
        behaviour. The system enters a reentrant nonmagnetic phase for $1.734 < J_1 < 1.872$ where $\Gamma_1$ is nonzero.
	The EE for the unsymmetrized GS is shown in (b), and the fidelity is shown in (c) as a function of $J_1$ for 
	the unsymmetrized GS.
        The EE exhibits a discontinuous change, while fidelity shows a sharp drop
        at the transition points. Wild fluctuations (many transitions) in both the EE and fidelity around $J_1=1.8$ and 2.3 are due to
	the degeneracy in the GS energy.}
\end{center}
\end{figure}
earlier studies it is established that the GS switches from a singlet to a triplet at $J_1 = 1.427$. For $1.734 < J_1 < 1.872$ the GS of 
the system again becomes a singlet and for $J_1 > 1.872$ the GS changes spin, finally 
attaining a $S_G = 3$ for $J_1 \ge 2.355$. In (Fig.~\ref{fig:nosym_all_57} (b) and (c)) we show how the EE and fidelity varies as a function 
of $J_1$. At the first transition, both these quantities show a sharp change. In the region of the reentrant phase, the variations are 
violent. In the region $ 1.872 < J_1 < 2.176$, EE shows smooth variation and the fidelity also does not change. In this region 
the GS spin, $S_G = 1$. Although the $S_G$ value continues to be 1 in the region $2.176 < J_1 < 2.355$, the EE as well 
as fidelity show very sharp changes. Finally beyond $J_1 = 2.355$, EE is almost constant and fidelity stays at 1.

The first transition from a singlet to a triplet is simple and the critical $J_1$ value can be pinned down both from EE and fidelity. This
occurs at ${J_1} = 1.427$. To understand the behavior at other values of $J_1$, we recognize that the system has a reflection symmetry
perpendicular to the legs. So, the states of the system can be classified as $\sigma {(+)}$ or $\sigma {(-)}$ depending on the symmetry of
the space in which the GS is found ~\cite{Geet}. The range of $J_1$ for the lowest energy state in different subspaces is shown in Table ~\ref{tab:57}.
\begingroup
\begin{table}
 \begin{center}
	 \caption{\label{tab:57} The range of $J_1$ over which the GS have different symmetries in a 5/7 skewed
	 ladder with $N = 24$ spins. The first column represents the range of $J_1$ for which the GS symmetry is 
	 $\sigma(+)$ and
	 similarly the second column represents the GS is $\sigma(-)$ subspace. The third
	 column gives the intervals of $J_1$ for which the GS is doubly degenerate.}
\noindent\hrulefill
\smallskip\noindent
\resizebox{\linewidth}{!}{%
\begin{tabular}{|l|l|l|}
\hline
\multicolumn{1}{|l|}{\quad \quad \quad $\sigma(+)$}
& \multicolumn{1}{|l|}{\quad \quad \quad $\sigma(-)$}
& \multicolumn{1}{|l|}{\quad \quad \quad $|\Gamma_{\sigma}|=0$} \\ \hline
\quad $1<J_1\le 1.427$ & $1.427\le J_1 \le 1.753$ & $1.753< J_1 \le 1.871$ \\
\quad \quad \quad and & \quad \quad \quad and & \quad \quad \quad and \\
$1.871<J_1\le 2.176$ & \qquad  $J_1>2.355$ & $2.176< J_1 \le 2.355$ \\ \hline
\end{tabular}}
\end{center}
\end{table}
We see that the change in the spin of the GS from $S_G = 0$ to $S_G = 1$ at $J_1 = 1.427$ is also accompanied by vanishing of $\vert\Gamma_{\sigma}\vert$.
EE shows a jump and fidelity shows a sharp dip at $J_1 = 1.615$ and 1.725. The change in EE and fidelity at 1.615 is
associated with the maximum in the $\vert\Gamma_{\sigma}\vert$ gap. Beyond $J_1=1.615$, the $\vert\Gamma_{\sigma}\vert$ gap begins to close and at $J_1=1.754$
the gap vanishes. In region between 1.753 and
1.871, $\vert\Gamma_{\sigma}\vert$ vanishes. In the unsymmetrized calculations both fidelity and entropy show sharp fluctuations, which are
suppressed when the degenerate ground states are obtained in the $\sigma {(+)}$ and $\sigma {(-)}$ subspaces (Fig.~\ref{fig:sym_all_57}).
In the $J_1$ region between 2.176 to 2.355 the GS is a spin triplet  and $\vert\Gamma_{\sigma}\vert$ also vanishes. Fidelity of the $\sigma {(+)}$ and $\sigma {(-)}$ show a
sharp dip at $J_1=2.176$ and $2.355$ respectively; the GS below 2.176 has $\sigma {(+)}$ symmetry and becomes degenerate with the lowest
state in $\sigma {(-)}$ symmetry in the region of $J_1$ ($2.176 < J_1 \le 2.355$) and for $J_1$ value above 2.355 the GS switches back to 
$\sigma {(-)}$ subspace. There are no sharp changes in either entropy or fidelity beyond $J_1 = 2.355$.
\begin{figure}[ht!]
\begin{center}
\includegraphics[width=2.9 in]{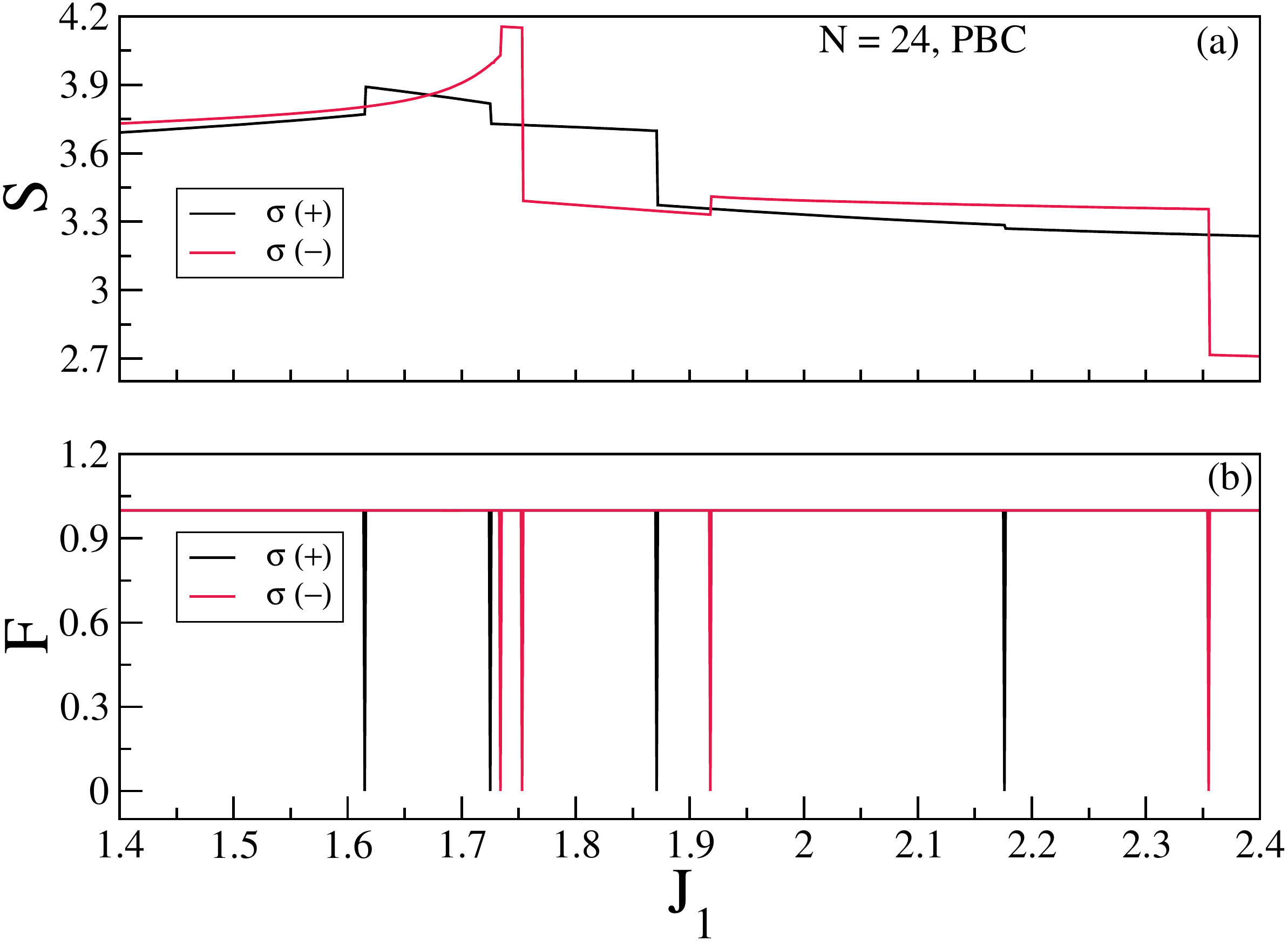}
        \caption{\label{fig:sym_all_57} For a 5/7 skewed ladder with 24 sites in PBC,
         (a) EE and (b) fidelity are plotted as a function of $J_1$ for both the reflection symmetry
         subspaces ($\sigma(+)$ and $\sigma(-)$).
         Black curve corresponds to the $\sigma(+)$ subspace, whereas the red curve represents the $\sigma(-)$ subspace.}
\end{center}
\end{figure}
\section{Summary}\label{sec5}
We have studied EE and fidelity to characterize the quantum phase transitions in skewed ladder systems. While correlation function also give
a measure of the extent of interactions between sites, EE embodies long range correlation between different parts of the system. In the
regions of $J_1$ where the system has a nondegenerate GS without change in spin or spatial symmetry the entropy change is gradual and
fidelity remains at one. At the transition points, the entropy shows a discontinuous change and fidelity shows a sharp dip. The transition
points are accurately determined from these characteristic changes. In regions where the GS is degenerate, the unsymmetrized calculations
show sharp fluctuations in EE and fidelity. However, in the symmetrized calculations they show abrupt changes only at the transition. We
also find that for regions where $\vert\Gamma_{\sigma}\vert$ vanishes, the fidelity does not change in either subspace. But at the beginning 
and at the end of the transition, fidelity shows sharp dip in the lowest eigenstate of one of the symmetry subspaces.\\
\section*{acknowledgments}
S. Ramasesha acknowledges the Indian National Science Academy and DST-SERB for supporting this work. Manoranjan Kumar acknowledges the SERB 
for financial support through Project File No. CRG/2020/000754.
\section*{Author contribution statement}
The project was conceived by S. Ramasesha and Manoranjan Kumar. Sambunath Das and Dayasindhu Dey have performed the numerical calculations. 
All the authors were involved in designing the calculations and interpretation of results. The work was jointly written up by all the authors.
\section*{Data Availability Statement}
The data that support the findings of this study are available from the corresponding author upon reasonable request.
\bibliography{bibliography}
\end{document}